\documentclass{aa}

\usepackage{graphicx}

\begin{document}

              
%
   \title{Linear polarization and composition of VLBI jets}

   \author{D. Fraix-Burnet
          }

   \institute{
Laboratoire d'Astrophysique de Grenoble,
BP 53, F-38041 Grenoble C\'edex 9, France.
             }
\mail{fraix@obs.ujf-grenoble.fr}

   \date{Received November 21 2000 / Accepted October 10 2001}

   \abstract{
It is shown that linear polarization data can be used to constrain the
composition (normal or pair plasma) of pc-scale extragalactic jets. 
 A simple criterion, based on synchrotron and Faraday depolarization
properties, is established. It does not depend on the particle density
and the length of the emitting region along the line of sight, thus
eliminating two physical unknowns.
      \keywords{galaxies : jet --
                galaxies : active --
                radiation mechanisms : non-thermal --
                radio continuum : galaxies --
                polarization --
                methods : analytical
              }
   }

   \maketitle

%

\section{Introduction}

Due to their synchrotron emission, extragalactic jets are known to
contain electrons. But what is the positive charge population : protons
or positrons? This makes a huge
difference in the kinetic energy to be put in the ejection process and
implies different properties for acceleration and propagation mechanisms. 

Extracted matter from the accretion disk is comprised of
normal (electron-proton) plasma. But accelerating protons to
relativistic speeds makes the energetic budget somewhat problematic. The scheme
of a two--fluid plasma (Sol et al. \cite{Sol}, Pelletier \& Marcowith \cite{pelletier}) alleviates this
difficulty by reserving the relativistic speed to a pair
(electron--positron) plasma, while the bulk of the jet, non relativistic
and non radiating, is made of normal plasma.

Observationally, there have been two approaches to determine the nature of the
radiating plasma: spectroscopic behavior
(Reynolds et al. \cite{Reynolds}), and circular polarization (Wardle et
al. \cite{Wardle}). The results tend very strongly toward
electron--positron pairs for the radiating component. 

In this paper, it is shown that linear polarization could also be used to
derive the nature of the jet material from very high resolution
observations. Based on basic
formulae presented in Sect.~2, a very simple analytical criterion is established
in Sect.~3. In Sect.~4, some particular observational data are
discussed before a brief conclusion is given in Sect.~5.

\section{Simple properties of synchrotron radiation}

Let us consider an electron population of density $N_{\mathrm{el}}$. The
positive charge is assumed to be a mixture of protons ($N_{\mathrm{pr}}$)
and positrons ($N_{\mathrm{po}}$). We have $N_{\mathrm{el}}=N_{\mathrm{po}}+
N_{\mathrm{pr}}$ and we define:

\begin{equation}
\tau = N_{\mathrm{po}} / N_{\mathrm{el}}.
\end{equation}

A simple homogeneous slab with a homogeneous magnetic field is assumed.
As explained later on, this assumption is not restrictive, but yields
the minimum depolarization that must be observed in a source. We only
consider the optically thin part of the synchrotron spectrum.

\subsection{Synchrotron surface brightness and polarization}

The synchrotron surface brightness of a population of electrons (or positrons) with an
energy distribution spectral index $p$ is derived from the standard
formulae (i.e. (5.43) in Ginzburg \cite{ginzburg}):

\begin{equation}
 I= K_1(p) 
\left(N_{\mathrm{el}}+N_{\mathrm{po}}\right)
\gamma_{\mathrm {min}}^{p-1} L 
B_{\perp}^{\left(p+1\right)/2}
{\delta}^{\left(p+n\right)/2}\nu^{\left(1-p\right)/2} 
\label{i}
\end{equation}
with
\begin{equation}
\begin{array}{ll}
 K_1(p) = 4.48~10^{-11}\left(\frac{p-1}{p+1}\right)
&\Gamma\left(\frac{3p-1}{12}\right)\Gamma\left(\frac{3p+19}{12}\right)
\\
&\;\;\;\;\;\;\times\left(8.40~10^{6}\right)^{\left(p-1\right)/2} 
\end{array}
\end{equation}
in ${\mathrm{ Jy~arcsec}}^{-2}$,
where $\Gamma(x)$ is the Gamma function, $N_{\mathrm{el}}+N_{\mathrm{po}}$ the total
radiating particle density, $\gamma_{\mathrm{min}}$ the minimum energy
of the electron (positron) population, $L$ the column length along the
line of sight in centimeters, $B_{\perp}$ the component of the magnetic
field which is perpendicular to the line of sight, $\nu$ the
observational frequency, $\delta$ the bulk Doppler factor, $n=3$ for
a continuous jet and $n=5$ for a single plasmoid (Lind \& Blandford \cite{LindBlandford}).

In the conditions assumed in this paper, the intrinsic polarization of the
synchrotron emission is given by (Ginzburg \cite{ginzburg}):

\begin{equation}
 \Pi_{\mathrm{i}} = \frac{p+1}{p+7/3} 
\end{equation}

which is $\simeq 69$\% for $p=2$ and $\simeq 75$\% for $p=3$.

\subsection{Faraday depolarization}

The polarized synchrotron radiation emitted by a given set of electrons
is affected by Faraday rotation through other sets of
electrons along the line of sight. Integration of the
light coming from all these sets results in an apparent depolarization
called the Faraday depolarization. This is the minimum depolarization
that occurs. Any heterogeneity (in density or in magnetic field) within
the slab increases the observed depolarization.

For positrons, the Faraday rotation is of opposite sign, so that there is
no Faraday depolarization in a pure electron--positron plasma. 

For protons, the Faraday rotation is negligible and does not compensate
for that of the electrons. Hence, the observed polarization
$\Pi_{\mathrm{obs}}$ is constrained by:

\begin{equation}
 \Pi_{\mathrm{obs}} \leq \Pi_{\mathrm{i}}\ \frac{\sin(\psi)}{\psi} \leq \Pi_{\mathrm{i}}\ \frac{1}{\psi} 
      \label{piobs}
\end{equation}

The Faraday rotation angle $\psi$ is given by (Burn \cite{Burn}):

\begin{equation}
 \psi = 2.35~10^4 \left(N_{\mathrm{el}}-N_{\mathrm{po}}\right)
B_{//}L\nu^{-2}\delta^2 
\label{psi}
\end{equation}

where $B_{//}$ is the component
of the magnetic field which is parallel to the line of sight.

\section{Combining surface brightness and polarization informations}

\subsection{Principle}

Both $I$ and $\psi$ are increasing functions of density, $B$ and $L$.
The brightest components should then be less polarized in the case of
a normal plasma. In principle, this could be an observational test for
determining the composition of a jet. However, this is only a statistical
argument, based on the improbable assumption that all parameters are equal,
particularly the level of homogeneity of the emitting region. Is there
another way of translating the incompatibility between high surface
brightness and high polarization for a normal plasma?

We propose to use the ratio between the surface brightness $I$ and the
Faraday depolarization $\psi$. As calculated in the
next section, its notable property is that it does not depend on
$N_{\mathrm{el}}$ and $L$ for a normal plasma, eliminating the
indetermination regarding these two parameters.

\subsection{Definition of the Brightness Faraday Ratio}

Using equations~(\ref{i}) and (\ref{psi}), we define
the ``Brightness Faraday Ratio'' as:

\begin{equation}
BFR =
 \frac{I}{\psi} = K_2(p) \epsilon \nu^{\left(5-p\right)/2}\gamma_{\mathrm{min}}^{p-1} 
\frac{B_{\perp}^{\left(p+1\right)/2} }{B_{//}}\delta^{\left(p+n-4\right)/2} 
%
\label{bfrdef}
\end{equation}
where
\begin{equation}
\begin{array}{ll}
 K_2(p) = K_1(p) / 2.35~10^4,
\end{array}
\end{equation}
in ${\mathrm{Jy~arcsec}}^{-2}{\mathrm{rad}}^{-1}$, and
\begin{equation}
\epsilon=\frac{\left(N_{\mathrm{el}}+N_{\mathrm{po}}\right)}
{\left(N_{\mathrm{el}}-N_{\mathrm{po}}\right)}=\frac{1+\tau}{1-\tau}.
\end{equation}

The function $\epsilon$
essentially $\approx 1$, except
for a nearly pure pair plasma in which case it can be much higher.
Let $v$ be the angle between the magnetic field and the line of sight.
It becomes:

\begin{equation}
\frac{B_{\perp}^{\left(p+1\right)/2} }{B_{//}} = B^{\left(p-1\right)/2} b(p,v)
\end{equation}
with
\begin{equation}
b(p,v) = \frac{\left(\sin v \right)^{\left(p+1\right)/2} }{\cos v}.
\end{equation}

   \begin{figure}
   \centering
   \includegraphics[angle=-90,width=5cm]{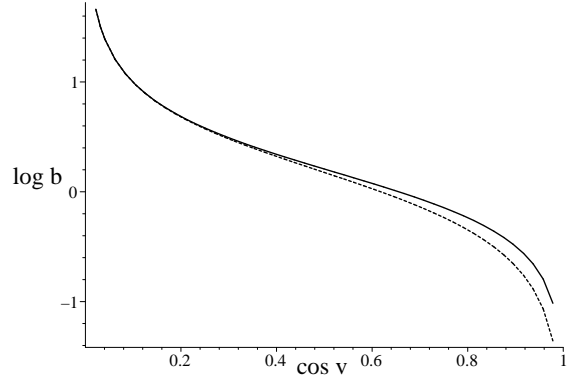}
   \caption{Function $log\left(b(p,v)\right)$ for $p=2$ (solid line) and $p=3$ (dotted line)
vs $\cos v$.}
              \label{bpv}
    \end{figure}

The function $b(p,v)$ is plotted on Fig.~\ref{bpv}. It is smaller than
0.1 for $v \leq 15\degr$
and is higher than 10 for $v \geq 85\degr$. 

Finally, $BFR$ can be written in the following way:

\begin{equation}
 BFR = K_2(p)\epsilon b(p,v) \nu^{\left(5-p\right)/2}
           \gamma_{\mathrm{min}}^{p-1} B^{\left(p-1\right)/2}\delta^{\left(p+n-4\right)/2}
\label{bfr2}
\end{equation}

\subsection{Numerical value for pc-scale jets}

The determination of the physical parameters in extragalactic jets is a
major problem. We now re-write
equation~(\ref{bfr2}) in a convenient form to numerically apply it to
jets.

As suggested by the observations, $p$ is somewhat between 2 and 3:
 $K_2(2)=0.103$ and $K_2(3)=0.074$. 
Even if it is a somewhat uncertain guess, a typical value for the
magnetic field appears to be $B\approx 10^{-2}$~G. $\gamma_{\mathrm{min}}$
seems to be more constrained at about $10^2$ (see discussion
in Reynolds et al. \cite{Reynolds}). The allowed
variations of these two parameters are probably not much higher
than a factor of $10$ in most cases.

Defining:

\begin{displaymath}
\nu_{10}=\left(\frac{\nu}{10^{10}\,{\mathrm{Hz}}}\right)\; ,\;
\gamma_{\mathrm{100}}=\left(\frac{\gamma_{\mathrm{min}}}{10^2}\right)\; ,\;
B_{-2}=\left(\frac{B}{10^{-2}\,{\mathrm G}}\right),
\end{displaymath}
we obtain:
\begin{equation}
\begin{array}{lll}
BFR_{p=2} =& 4.06~10^{4}\epsilon b(2,v) \delta^{\left(n-2\right)/2}
           \nu_{10}^{3/2} &\gamma_{\mathrm{100}}
           B_{-2}^{1/2}, \\[0.3cm]
BFR_{p=3} =& 1.29~10^{4}\epsilon b(3,v) \delta^{\left(n-1\right)/2}
           \nu_{10}&\gamma_{\mathrm{100}}^2 B_{-2}.\\
\label{bfrnum}
\end{array}
\end{equation}

Since $BFR$ is intended to test the normal plasma hypothesis, one can
assume $\epsilon\approx 1$ (see Sect.~3.2).

The value for $n$ depends on the geometry of the source (see Sect.~2.1). However, jets
are characterized by a complex continuous structure so that most
probably: $n=3$. This makes a relatively softer dependance on the Doppler
factor than for $n=5$.

This Doppler factor is known from apparent motions in quite a lot of jets. 
It seems to be rarely higher than about 10 (i.e. Zensus \cite{Zensus}).  

The most uncertain parameter in the expression of $BFR$ is undoubtly the
angle $v$ which cannot be easily determined from the observations. However, as said
previously (see Fig.~\ref{bpv}), $b(p,v)$ is higher than 10 only when
$v$ is larger than $85\degr$ (and $b(p,v)>5$ when $v>80\degr$), that is 
when the magnetic field is nearly exactly in the plane of the sky. 
In practice, by
comparing a given set of several structures within a jet and/or in different sources,
it would certainly be possible to restrain $v$ to a reasonable range
where the function $b(p,v)$ can be given a representative upper limit. 

Finally, the choice of $p$ between 2 or 3 might not be so critical.

\subsection{A criterion for the composition of jets}

For a given set of physical parameters, $BFR$ is the maximum value allowed for $I/\psi$
in the case of
an electron--proton plasma. Combining equations~(\ref{piobs}) and 
(\ref{bfrdef}) the scheme is therefore very simple:

\begin{equation}
\frac{I_{\mathrm{obs}}\Pi_{\mathrm{obs}}}{\Pi_{\mathrm i}} > BFR \Rightarrow \mathrm{normal\ plasma\ excluded}
\label{applic}
\end{equation}
where $I_{\mathrm{obs}}$ is the observed surface brightness.

To avoid the contribution of smearing in the observing beam that can
particularly affect the observed polarization and also the intensity, the highest
spatial resolution is recommended. As a consequence, the VLBI
observations are expected to provide the most interesting data.

Note that criterion~(\ref{applic}) can be used with other criteria like
those in Reynolds et al. (\cite{Reynolds}) to reduce uncertainties on parameters.

\section{Comparison with observations}

In this section, we illustrate the use of the $BFR$ criterion with observations.
A concrete application of the proposed
criterion requires a thorough study of individual cases or of a
statistically significant sample, in order to determine the most appropriate values for
the parameters in $BFR$.

\subsection{\object{M87}}

We start with the example of knot A in the jet of \object{M87} at
0.14~arcsec resolution and 15~GHz as observed by Owen et al. (\cite{owen}). The
surface brightness from their map is found to be about 15~Jy
arcsec$^{-2}$ with a $\approx 30$\% polarization. This yields:

\begin{displaymath}
\frac{I_{\mathrm{obs}}\Pi_{\mathrm{obs}}}{\Pi_{\mathrm i}}\approx 6.5
\end{displaymath}

In this case, it is not possible to draw conclusions about the composition of
the jet, but at this resolution, beam smearing is quite certain and the
use of the criterion is somewhat hopeless. Taking the observations by
Junor et al. (\cite{Junor}) at 43~GHz and 0.33~mas$\times$0.12~mas resolution, we
obtain a surface brightness of about $10^6$~Jy arcsec$^{-2}$. Assuming a
polarization of 10\%, this would give
$I_{\mathrm{obs}}\Pi_{\mathrm{obs}}/\Pi_{\mathrm i} \approx 1.4~10^5$.
Preserving a normal plasma hypothesis would put some interesting
contraints on the parameters in equation~(\ref{bfrnum}).
This illustrates the usefulness of
the criterion presented in this paper when applied to very high
resolution data.

\subsection{\object{3C273}}

In Lister \& Smith (\cite{ListerSmith}), both
surface brightness and polarization of several components in different sources
are compiled.
For component G of the jet in \object{3C273} we derive a surface brightness of
about $5~10^5$~Jy arcsec$^{-2}$ and a rather low polarization of 4\%.
However, this still yields at 22~GHz (resolution of 0.97~mas$\times$0.38~mas):

\begin{displaymath}
\frac{I_{\mathrm{obs}}\Pi_{\mathrm{obs}}}{\Pi_{\mathrm i}}\approx
3~10^6.
\end{displaymath}

In components C and D, we have a surface brightness of about $4~10^5$~Jy
arcsec$^{-2}$ with about 21\% polarization. This yields:

\begin{displaymath}
\frac{I_{\mathrm{obs}}\Pi_{\mathrm{obs}}}{\Pi_{\mathrm i}}\approx
10^5.
\end{displaymath}

These high observed ratios require, for a normal plasma,
high values of the parameters (i.e. $n=5$, large Doppler factor, high
value for $b(p,v)$...). 

One way to reduce some uncertainties on the parameters would be to
compute the above ratio in every structure of the jet. For instance, it
would be hard to believe that $v$ is the same everywhere in the jet,
hence it would be possible to put a global upper limit on $b(p,v)$. Note
also that the composition of the radiating material is certainly the
same everywhere in a given jet.

\subsection{\object{3C279}}

From the same paper (Lister \& Smith \cite{ListerSmith}), we find 
that component F in the jet of \object{3C279}, at about the same
resolution, has
a surface brightness of about $1.4~10^7$~Jy arcsec$^{-2}$ and 18\%
polarization. This yields:

\begin{displaymath}
\frac{I_{\mathrm{obs}}\Pi_{\mathrm{obs}}}{\Pi_{\mathrm i}}\approx
3.6~10^6.
\end{displaymath}

This again shows that our criteria provide interesting
constraints on the parameters in equation~(\ref{bfrnum}) if one wishes to
assume a normal plasma. The fact that the ratio found in knot F of 3C279
is close to that for knot G in 3C273 is undoubtly pure chance, but
comparing values found in several jets is another way to use the
$BFR$ criterion and find hints for the composition of jets in general.

\section{Conclusion}

The present study has been initiated from the synchrotron emission
simulations (Despringre \& Fraix-Burnet \cite{paper1}, Fraix-Burnet in
prep.) in which it appears nearly impossible to reproduce both surface
brightness and polarization of typical VLBI jets. We have then devised a
simple tool, presented in this paper, to constrain the composition of a
jet with linear polarization data. 

It is based on the Faraday depolarization which is the minimum
depolarization that occurs from a normal plasma. This yields a
criterion, easily computed from the observations, that does not depend on
the length of the emitting plasma along the line of sight nor on density
of the emitting particles. These two parameters are always very
difficult to estimate. 

Still, there are a few uncertain parameters in the criterion, but from a few
illustrative examples presented in this paper, it already appears that
they should be pushed to high values if one assumes a normal plasma. 
A complete statistical study of several jets or a very detailed study
of a well-observed jet is necessary to reach more significant
conclusions. This is beyond the scope of the present paper.

Our criterion is not valid in the
self--absorbed part of the spectrum. It is certainly not a problem for
frequencies above 10~GHz in the jets, but it should be applied to cores only
with caution.



\end{document}